\documentclass [12pt,twoside]{article}           
\usepackage[left,modulo]{lineno}
\usepackage{setspace}

\countdef\arxiv=201

\ifnum\arxiv=0 \input cpreb.tex \fi

\ifnum\arxiv=1

\usepackage{epsfig,times,lscape}
\usepackage[usenames]{color}

    \ifnum\count102=1

    \fi

    \ifnum\count102=2
\fi

\newcommand{\nc}{\newcommand}

     \ifnum\count101=1
\nc{\qI}[1]{\section{{#1}}}
\nc{\qA}[1]{\subsection{{#1}}}
\nc{\qun}[1]{\subsubsection{{#1}}}
\nc{\qa}[1]{\paragraph{{#1}}}

\def\qpar{\vskip 2mm plus 0.2mm minus 0.2mm}
\def\qL{\hfill \break}
     \fi 

      \ifnum\count101=2
 \nc{\qI}[1]{\parindent=0mm \vskip 8mm 
{\centerline{\LARGE \color{red}#1}}\vskip 3mm}
\nc{\qA}[1]{\vskip 2.5mm \noindent 
{{\bf\large\color{blue}  #1}} \vskip 1mm \parindent=0mm}
 \nc{\qun}[1]{\vskip 1mm \noindent {\sl #1 }\quad }

\def\qL{\hfill \break}
\def\qpar{\vskip 2mm plus 0.2mm minus 0.2mm}

      \fi

\nc{\qfoot}[1]{\footnote{{#1}}}

      \ifnum\count101=1
\def\qbu{\hfill \par \hskip 6mm $ \bullet $ \hskip 2mm}
\def\qee#1{\hfill \par \hskip 6mm (#1) \hskip 2 mm}
      \fi
      \ifnum\count101=2
\def\qbu{\hfill \par \hskip 4mm $ \bullet $ \hskip 2mm}
\def\qee#1{\hfill \par \hskip 4mm (#1) \hskip 2 mm}
      \fi

\def\qparr{ \vskip 1.0mm plus 0.2mm minus 0.2mm \hangindent=10mm
\hangafter=1}

     \ifnum\count101=1 
 \def\qdec#1{\parindent=0mm\par {\leftskip=2cm {#1} \par}}
     \fi
     \ifnum\count101=2
  \def\qdec#1{\parindent=0mm \par {\leftskip=1cm {#1} \par}}
  
  \def\qcitb#1{\noindent \hbox to 102mm{\hfill \small #1} \vskip 1mm}
      \fi

 \def\qpages#1{\count102=0{\loop\advance\count102 by 1
 \null \vfill\eject \ifnum\count102<#1 \repeat}}

\def\qn#1{\eqno \hbox{(#1)}}

\def\qv{\vskip 0.1mm plus 0.05mm minus 0.05mm}

\def\qhv{\hskip 3mm}

\def\qhw{\hskip 1.5mm}
\def\qleg#1#2#3{\noindent {\bf \small #1\qhw}{\small #2\qhw}{\it \small #3}\qv }
\fi

\begin{document}

\thispagestyle{empty}
\null\vskip 2cm

\centerline{\Large \bf \color{blue} Spatial and historical determinants}
\vskip 0.5cm 
\centerline{\Large \bf \color{blue} of}
\vskip 0.5cm 
\centerline{\Large \bf \color{blue} separatism and integration}
\vskip 0.5cm 
\centerline{\Large \bf \color{blue} 2. Quantitative analysis}
  
\vskip 0.5cm

\centerline{\large  Bertrand M. Roehner$^1$}
\qpar
\centerline{\large Institute for theoretical
and High Energy Physics, University of Paris 6}

\large
\vskip 1cm

{\bf Abstract}\quad In part 1 we have discussed the role
played by geographical and historical conditions.
Here we wish to test our model and to estimate its
parameters. Ti this effect
two indexes are introduced. The geographical index
is aimed at characterizing the degree of separateness of a
given region with respect to the national state to which it
belongs. This index turns out to be closely connected to the
number of minority-speakers; it is also
correlated with the level of separatist disturbances, at least for
samples having the same historical background. The purpose of the
second index is precisely to specify the role of past episodes in
shaping current separatist outbreaks. To a large extent,
current episodes are  modeled on former ones; typically
this process of semi-replication appears to be characterized by a
``memory'' that extends over at least one and a half century.
Note that this study concentrates on the occurrence and forms of
separatist struggles rather than on their 
short-term political causes. 
In so doing, it
follows a methodological track that has been pioneered by 
Stanley Lieberson
(1985, the ``stardom'' paradigm) and Charles Tilly 
(1993, the ``traffic jam'' paradigm).

\vskip 0.5cm
\centerline{5 June 2017. Comments are welcome.}
\vskip 0.5cm

{\normalsize
{{1: } Postal address: LPTHE,
University Paris 6 (Pierre and Marie Curie), 4 Place Jussieu, 75005
Paris   \qL
\phantom{1: } Email address: roehner@lpthe.jussieu.fr
\qL 
\phantom{1: } Phone: 33 1 44 27 39 16}
}

\vfill \eject

\qI{Introduction}

In the first part of this study, thereafter referred to as paper
1, we have discussed the relevance of geographical and historical
factors in the apparition and long term evolution of separatist
movements. In the present paper we examine how and to what extent
our previous qualitative discussion can be substantiated by
quantitative evidence. This first requires the model delineated
in paper I to be defined more precisely. In section 2 we
define a geographical index that describes the degree of
separateness of a given region from the rest of the country. We
also introduce a historical index aimed at summarizing the
region's historical record. In section 3 we analyze 
mother-tongue maintenance for minority languages. It will be seen
that spatial factors are major determinants as
has already been emphasized in a different context by Lieberson
(1975). Next we try to assess the respective role of geographical
and historical factors in a number of separatist movements.
Finally, we point out that in spite of the current vigor of
separatist tendencies, a sustained process of economic
centralization is currently under way in most countries; this
process seems to be largely independent of the country's political
structure (i.e. federalist versus centralized). 
\qpar

Before we proceed a word of caution may be in order. 
In making our discussion quantitative, we shall
concentrate our attention on a small number of basic factors for we
do not really believe in the explanatory power of high dimensional
multivariate correlation analysis. 
 
\qI{Definition of the spatial and historical indexes}

\qA{How to characterize spatial separateness?}

In paper I we used the length of the frontier between a given
minority region $ (a) $ and the rest of the country $ (B) $ as a
way to assess the strength of the interaction between $ a $ and $
B $. We also emphasized that, should there be a contact zone with
another ``$ a $-speaking area'', its contribution should be
subtracted%
\qfoot{An illustrative example is provided by the Acadians
established in New Brunswick; it is certainly easier for them to
maintain their French mother tongue than it is for those living in
other Maritime Provinces; one only needs to mention the following
circumstances: possible contacts with family members who emigrated
to Quebec, availability of local French newspapers,
opportunity to watch TV programs from Quebec, etc.}%
.
For the purpose of quantitative tests this definition has to be
refined however. In particular, it will be convenient to use a
normalized index whose magnitude will be independent of the
size of the region under consideration. To this end we introduce the
following definition:
\qpar
\qdec{{\bf Definition 1}\quad We consider a region $ (a) $ where
a minority language is spoken. Let us introduce the following
notations:
\qbu $ l_0 $: \quad length of the coast-line of $ a $
\qbu $ l_+ $: \quad length of the contact zone of $ a $ with the
rest of the country $ (B) $.
\qbu $ l_- $: \quad length of the borderline (if any) of $ a $
with another $ a  $-speaking area $ (A) $ belonging to a foreign
country. Then the index of spatial separateness for region $ a $ is
defined as:
$$ g = {l_+w_+ - l_-w_- \over l_0 + l_+w_+ + l_-w_-} \qn{1.1} $$
$ w_+ $ (respectively $ w_- $) is a parameter characterizing the
intensity and frequency of the contacts of region $ a $ with $ B $ 
(respectively $ A $).\qL
$ g $ is normalized in the sense that it is comprised between $ -1
$ and $ 1 $. 
}

\qA{Examples}

  \begin{figure}[tb]
    \centerline{\psfig{width=7cm,figure=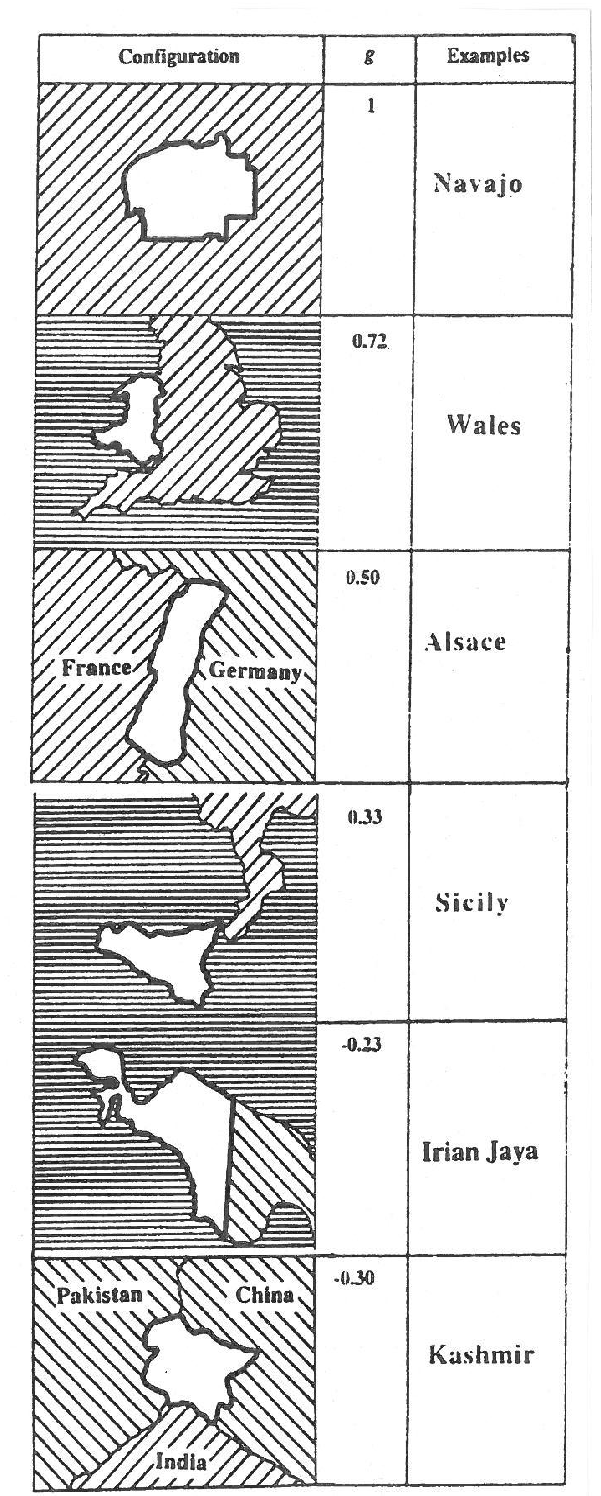}}
\qleg{Fig. 1\qhv  Examples of regions with minorities.}
{The homelands of the minorities are represented without cross-hatching;
the regions where the official
language of the state is spoken are represented by ascending cross-hatching;
the regions where the minority language is spoken are represented 
by descending cross-hatching.
The maps
in the left-hand column correspond to the cases mentioned in the
right-hand column; the middle column gives the corresponding values of the
geographical index of interaction; the above
cases have been classified by decreasing
order of the index (remember that the values of this index are 
comprised between -1 and 1).}
{}    
 \end{figure}

It may be enlightening to apply the above definition to a few
examples. The following are illustrative examples (Fig. 1)
$$ \matrix{
\bullet \quad a = & \hbox{Wales,} \hfill & B= & \hbox{Britain} & :
&  g= 0.52 \hfill \cr  
\bullet \quad a = & \hbox{Alsace,} \hfill & B= &
\hbox{France} & : &  g= 0.50 \hfill \cr
\bullet \quad a = & \hbox{Kashmir,} \hfill & B= & \hbox{India} & :
& g \simeq -0.30. \hfill \cr
} $$

In paper I we
gave a simple scale for the estimation of $ w $; let us recall it
briefly:
\qbu $ w=1 $: for ``traditional'' societies
\qbu $ w=2 $: for intermediate rural societies
\qbu $ w=3 $: for industrialized societies
\qpar

So far we did not examine to what extent $ w $ is affected by the
presence of a national borderline. Let us discuss the matter on
the example of Alsace (France); in spite of the close links
between Alsace and Baden in Germany, it is clear that $ w_+ $
(contacts with France) is markedly greater than $ w_- $ (contacts
with Germany) if only for the obvious reason that French is used
as the principal language both in school and in the
administration; while $ w_+ $ is clearly equal to $ 3 $ the value
that should be attributed to $ w_- $ is less obvious; indeed  $ w_-
$ is much more time dependent than is $ w_+ $: the number of
Alsatians working in Germany, the attractiveness of German TV
programs, and many other similar conditions are likely to change
fairly rapidly in the course of time. Given all these uncertainties
we selected $ w_- =1 $ as being a simple and not quite unreasonable
figure.  \qpar

{\bf The case of islands} \quad So far, we did not consider the
case of islands. Our former definition would lead to $ g = 0/l_0 =0 $.
The value   $ g=0 $
is clearly too crude. The matter is discussed in some
detail in Appendix C. It leads us to the following definition.
\qpar

\qdec{{\bf Definition 2}\quad We consider an island $ (a) $ where
a minority language is spoken. Let us introduce the following
notations:
\qbu $ d_+ $: average distance between the ports of $ a $ and those
of the mainland $ (B) $.
\qbu $ d_- $: average distance between the ports of $ a $ and
those of an $ a $-speaking region $ A $ belonging to a foreign
country.
\qbu $ w_+ $ and $ w_- $ have the same meaning as in Definition
1. \qL
The index of spatial separateness for island $ a $ is defined
as:
$$ g = { 1 \over D } { w_+/d_+ - w_-/d_- \over w_+/d_+ + w_-/d_- }
$$
where $ D $ is a normalization factor expressed in hundreds of
kilometers as:
$$ D= \hbox{Max}[\hbox{Inf}(d_+,d_-),3] $$
$ g $ is normalized in the sense that it is comprised between $
-0.33 $ and $ 0.33 $.
}
\qpar

As a matter of illustration, we apply the above definition to the
example of the Caribbean island of Guadeloupe which is a French
territory; in this case $ d_+ = 7,000 $ km; the nearest territory
having Creole as its official language is Haiti: $ d_- = 500 $ km;
with $ w_+ = 3,\ w_- = 1 $ (see Appendix C) we obtain: $ D=d_- = 5 $
and $ g = 0.05 $. 
\qpar

{\bf Remark}\quad As explained in Appendix C, one of the purpose
fulfilled by definition 2 is to guaranty a smooth transition
between the case of a peninsula (which is ruled  by definition 1)
and the case of an island. This can only be achieved in an
approximate way however; indeed it is clear that coastal
navigation plays a substantial role in the case of a peninsula, a
factor which has been ignored in definition 1. Such inaccuracies
should not be taken too seriously however for the main uncertainty
stems in fact from the difficulty of obtaining reliable estimates for
the mobilization parameter $ w $.

\qA{A quantitative statement of the paronymy hypothesis}

In paper I the paronymy assumption has been introduced, formalized
and to some extent confronted to empirical evidence. As may be
remembered, we observed a high degree of historical continuity in
the manifestations of separatist feelings. For the purpose
of making paronymy into a quantitative parameter
we introduce an index aimed at assessing the intensity of
separatist struggles in the following way.
\qpar

\qdec{{\bf Definition 3}\quad We consider a region $ a $ which is
the homeland of a minority; its resistance to assimilation is
characterized by an index $ h $ defined in the following way:
\qbu $ h = 0 $, if the region has not been annexed
\qbu $ h = 1 $, if there has been a peaceful annexation and no
subsequent separatist claims.
\qbu If there have been separatist disturbances, we take for $ h $
the logarithm of the total number of deaths. 
}

\qI{Estimating relaxation times for minority languages}

\qA{The evolution in the number of minority-language speakers}
In the life of a language%
\qfoot{In principle one should  make a distinction between various
ability levels. Four stages may be considered depending on whether
a language is understand, spoken, read or written. In the
case of minority languages statistics depend drastically on
which stage one considers. For instance in the Basque Provinces
(Spain) the respective percentages are: 50,46,25 and 11;
in Catalonia (Spain) they are: 90,77,62 and 38 (Minority Rights
Group 1977). Unfortunately,  only the number of speakers is in
general recorded and usually with a fairly high margin of error
(see in this respect the discussion in Kirk 1946, p.224.}%
, 
two phases should be distinguished: the
equilibrium phase and the assimilation phase.
%
  \begin{figure}[tb]
    \centerline{\psfig{width=15cm,figure=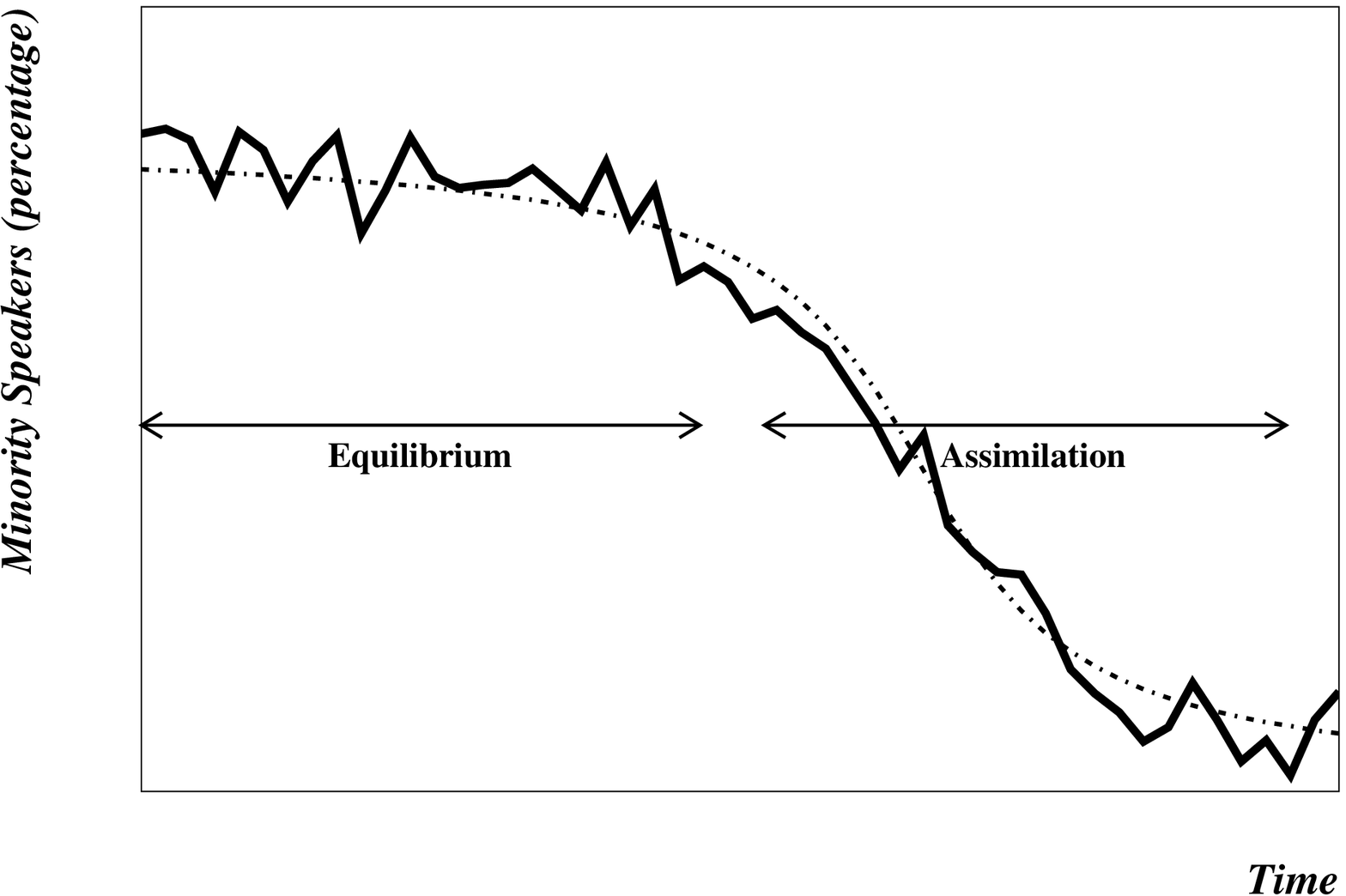}}
\qleg{Fig. 2\qhv Schematic Representation of the two 
phases in the life of a language. Schematic representation of the two 
phases in the life of a language.}
{The equilibrium phase
corresponds to average stability in the number of speakers. The
assimilation rate is represented by the slope of the curve. In the
equilibrium phase the assimilation rate is almost equal to zero. During the
transition to the fall-off phase, the assimilation rate gradually increases
until the largest part of the population has been assimilated.}
{}    
 \end{figure}
 
Fig. 2 schematically
represents a typical evolution. Notice that both
phases may last for a very long time. In principle one should also
consider the possibility that the decline may be checked with the
subsequent occurrence of a revival phase. Such cases are very
rare however. The so-called linguistic revival that occurred in
the late nineteenth century in a number of countries (Albania,
Bielorussia, Flanders, Hungary, Iceland, Norway, Poland, Ukraine,
Uzbekistan, etc) were of a very different nature for in such cases
the  language was still spoken by the people as a dialect or more
precisely as a variety of dialects. The revival rather concerned the
recognition of that dialect as a proper language with its own
literary traditions. In contrast, the possible revival that
we consider here refers to the total number of speakers; to our
knowledge only one revival of that kind may currently be under way,
namely that of the Welsh language; still it has to be confirmed by
further statistics in coming decades.
\qpar

As a first approximation, the decline phase for language $ a $ may
be described by the following equation:
$$ p = p_0 e^{-at} = p_0 e^{-t/\tau} \qn{3.1} $$

where $ p $ denotes the percentage of $ a $-speaking people.  
The coefficient $ a $ may be referred to as the assimilation rate;
its inverse $ \tau = 1/a $ has a simple interpretation: in the
time interval $ \tau $ the percentage $ p $ is divided by a factor
2.7; $ \tau $ is called the relaxation time of the decrease phase.
The quasi-equilibrium situation which is schematized by the
left-hand part of the curve in Fig.2 corresponds to a very small
assimilation rate, i.e. to a very large relaxation time.

\qA{Relationship between the assimilation rate and the
geographical index $ g $}

The relation that we expect to hold between the assimilation rate
$ a $ and the geographical index $ g $ is depicted graphically in
Fig.3. 
  \begin{figure}[tb]
    \centerline{\psfig{width=15cm,figure=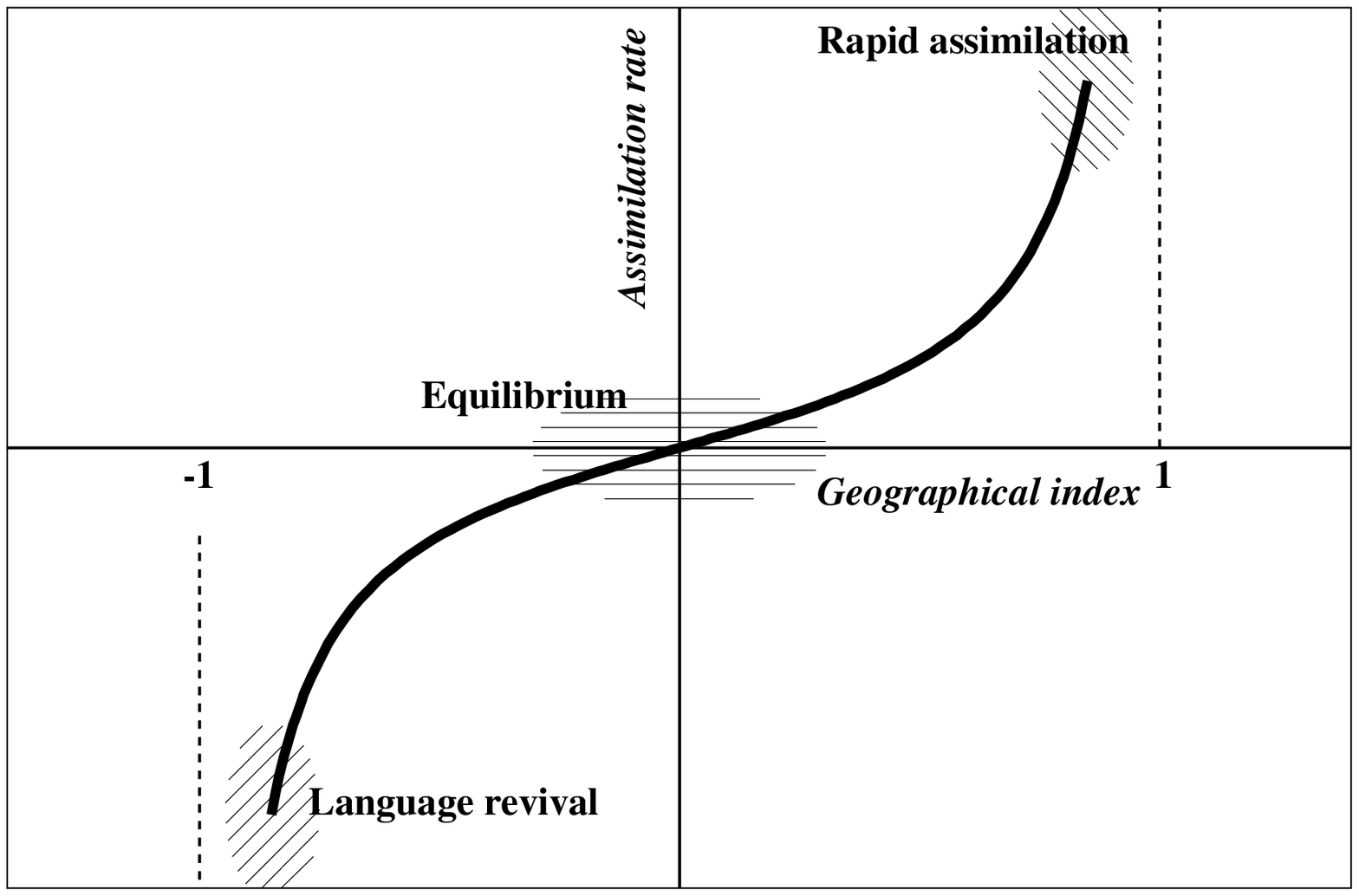}}
\qleg{Fig. 3\qhv Expected Relationship between 
the Geographical Index $ (g) $ and the Assimilation Rate $ (a) $.}
{The geographical index is close to 1 when the region
of the minority speakers is surrounded on all sides by majority speakers;
it is close to -1 when that region is better connected to (alien) minority
speakers than to majority speakers. In the first case assimilation occurs
fairly quickly (that is to say within two or three generations); in the
second, on the contrary, assimilation is a very problematic outcome.}
{}    
 \end{figure}
The following special cases are of particular significance:
\qee{1} $ g = 1 $: Rapid assimilation; example: the
``melting-pot'' period in the United States. 
\qee{2} $ g = 0 $: Equilibrium; examples: French-speaking Swiss or
Catalan-speakers in Spain. 
\qee{3} $ g = -1 $: Strong language revival; a rather hypothetical
situation as we already noticed. 
\qpar
  \begin{figure}[tb]
    \centerline{\psfig{width=13cm,figure=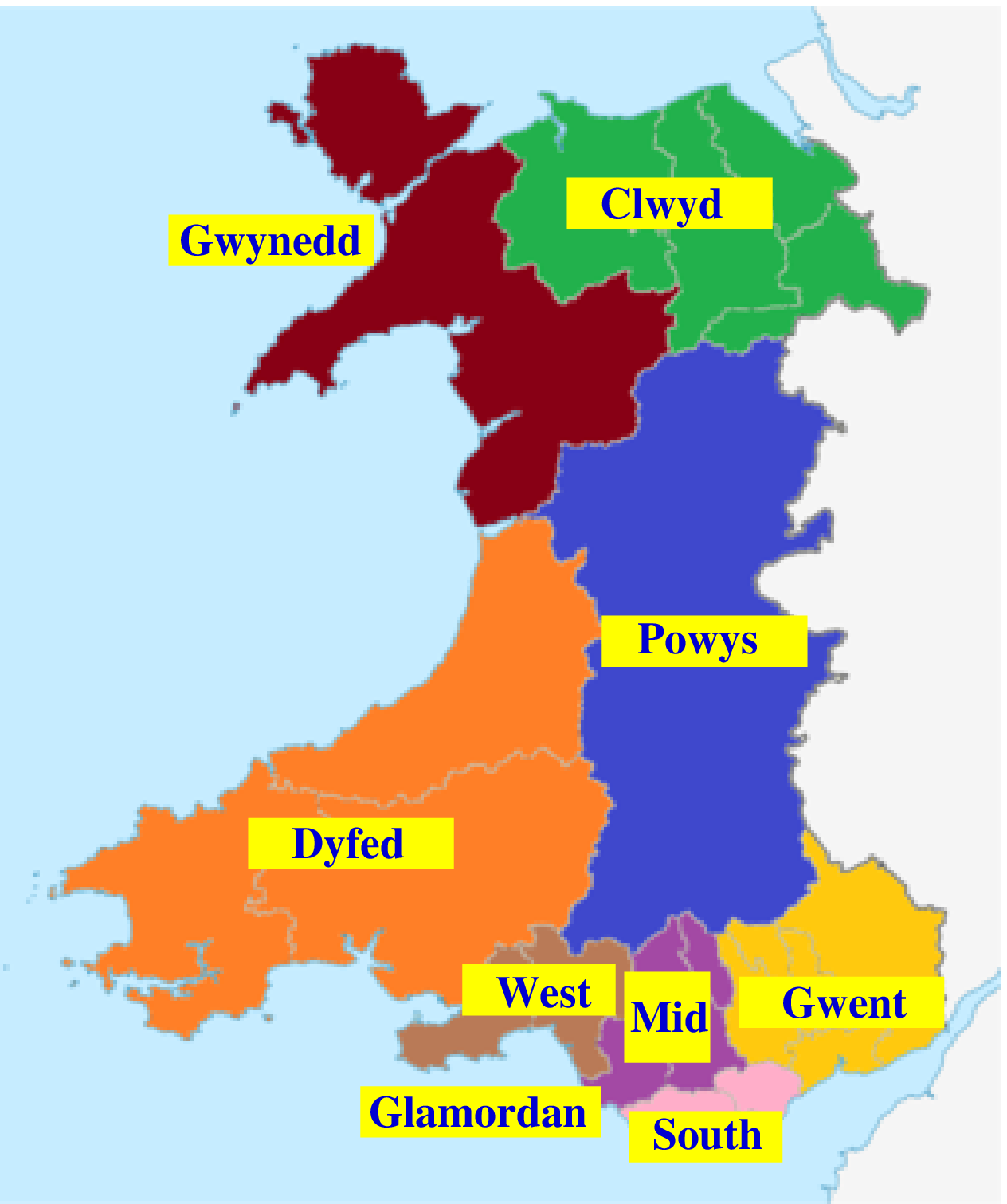}}
\qleg{Fig. 4a\qhv The counties of Wales. }
{The Welsh language is more spoken in the counties
which are far away from the border with England,
that is to say Dyfed and Gwynedd.}
{}    
 \end{figure}
%
  \begin{figure}[tb]
    \centerline{\psfig{width=15cm,figure=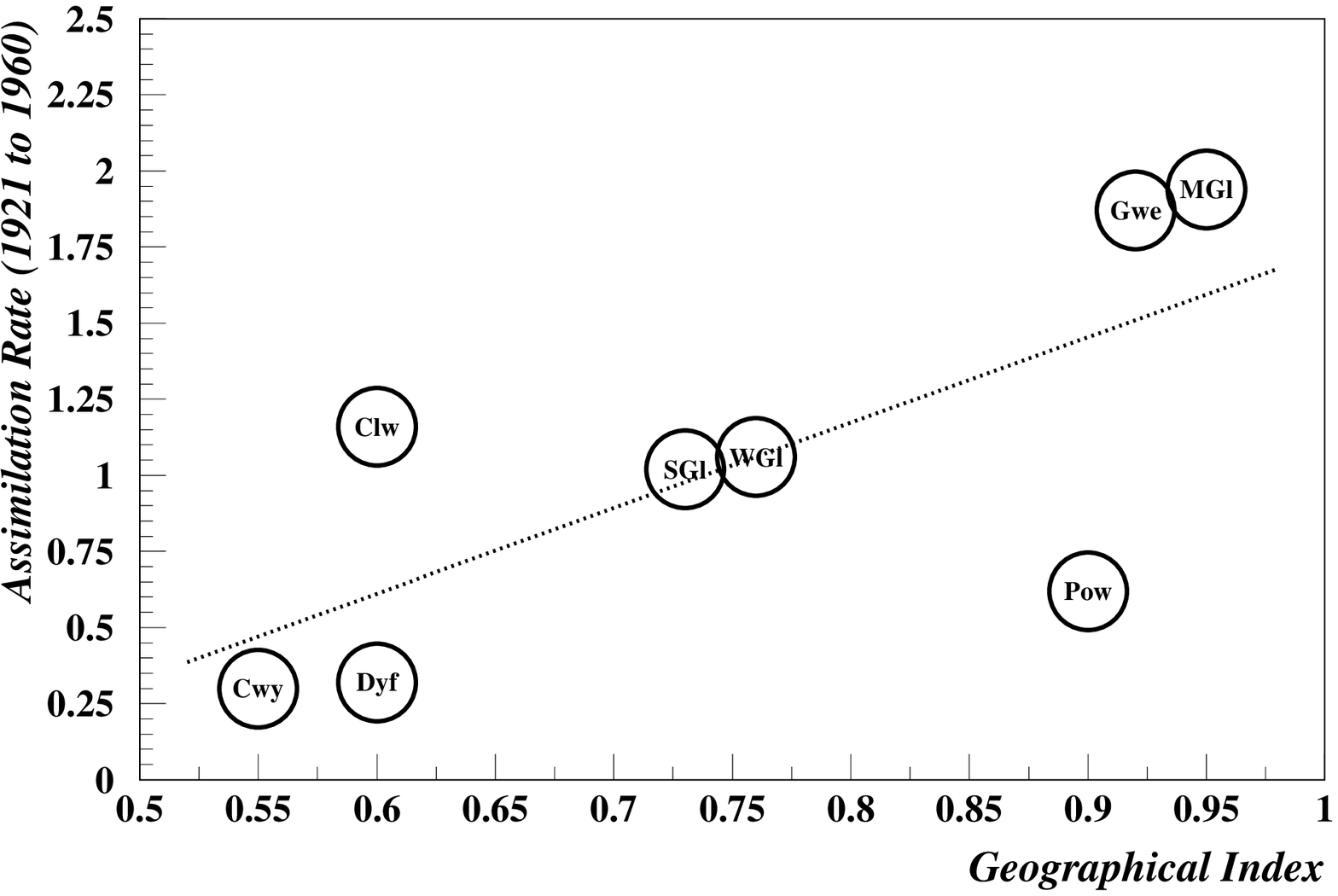}}
\qleg{Fig. 4b\qhv Observed Correlation between the geographical 
index $ (g) $ and the assimilation rate $ (a) $ in Wales.}
{The geographical index has been computed on the
following basis: counties with a number of speakers over 50 percent were
considered Welsh (only Gwynedd fell in this category), whereas the other
counties were considered as English. The correlation is equal to 0.71
(confidence interval at probability 0.95 is: 0.01 to 0.94).
Signification of the symbols  for the
counties in Wales: Clw: Clwyd; Dyf: Dyfed; Gwe: Gwent; 
Cwy: Gwynedd; MGl: Mid Glamorgan; Pow: Powis; SGl: South Glamorgan;  
WGla: West Glamorgan.}
{Source: See table 4.4 in Roehner (2002)}    
 \end{figure}
%
  \begin{figure}[tb]
    \centerline{\psfig{width=15cm,figure=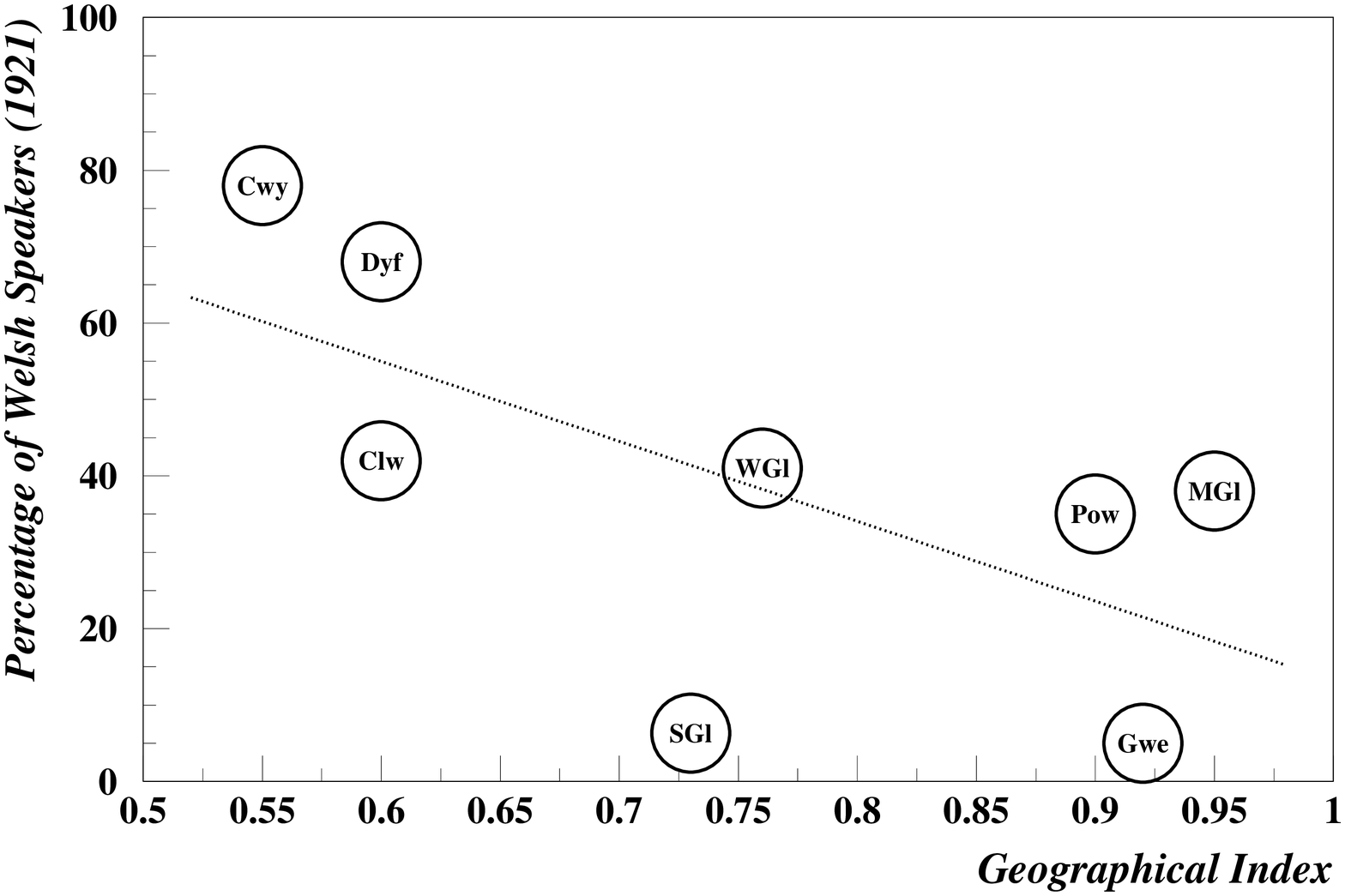}}
\qleg{Fig. 4c\qhv Observed Correlation between the geographical 
index $ (g) $ and the percentage of Welsh speakers in 1921.}
{The latter can be considered as
providing an estimate for the assimilation rate in the century before 1921.
The correlation is equal to -0.65 (confidence interval at probability
0.95 is: -0.92. Signification of symbols: see previous figure
to 0.10).}
{Source: Same as previous figure.}    
 \end{figure}

It can be noted that the rate of assimilation is largely
independent of the state's policy. Thus, despite of the fact that
Romantsch has become a national language (but not an official
language) in Switzerland after the Second World War, its decline
not only continued but even accelerated. The assimilation rate,
moreover, is of the same order of magnitude or even higher than it
is for minority languages in more centralized countries such as
Britain or France. While Romantsch was retreating, French and
Italian preserved their positions in Switzerland; in other words,
the potential usefulness of the language and the geographical
environment seem to be more important than linguistic policies.

\qI{The analysis of separatism}

\qA{The data}

Separatism usually reflects many causes, whether linguistic,
religious, cultural, ethnic or economic. One could even argue that
in joining a separatist struggle each individual has his own
motivations. As a result, it is almost impossible at the level of
an outside observer to distinguish between various sources of
separatism. Our methodological choice therefore was to estimate
the {\it overall} intensity of separatist struggles. To this end
we converted all major events into a number of fatal casualties.
In developing countries where separatist struggles often take the
form of open warfare, this number is almost identical with the
estimated number of deaths. On the contrary, in industrialized
societies it provides an equivalent figure for the demonstrations,
arson attemps, bomb attacks and other forms of protest. Specific
details about equivalence factors are given in Appendix A. 

\qA{The ceteris paribus condition}
 In a sense
the fact that almost all separatist movements are taking place in
long-standing enclaves is a first proof enough of the significance
of spatial determinants; yet, in order to make the test more
quantitative one has to make sure that variations in other factors
shall not spoil the test. 
In other words, in trying to assess the impact of the geographical
factor $ g $ we are confronted with the ceteris paribus requirement.
The following precautions are of
crucial importance.  
\qbu One should compare separatist movements
which are more or less at the same stage. It would be pointless to
compare a situation where national awakening is just under way (as
in Peru or Mexico for instance) with another where national
conscienceness is fully developed (as in Punjab for instance). 
\qbu The type of the polity has an obvious influence on the level
and forms of separatist disturbances. The latter do not take the
same form in a nation that has a long democratic tradition or in
one that has not. The above qualifications are far less crucial in
analyzing the impact of the historical factor for, in a sense,
the magnitude of the historical index reflects (though for an
earlier period) both the nature (more or less violent) of the
nationalistic movement and the response (more or less tolerant)
of the state. \qL
Owing to the above arguments, we shall carry out the following
tests:
\qee{1} The impact of the geographical factor $ g $ (taken alone)
is analyzed for separatist struggles in France: 9 cases. 
\qee{2} The impact of the historical factor $ h $ is considered
separately first for ``old'' (i.e. European)  countries, and then
for ``new'' countries: 6 and 9 cases respectively. 
\qee{3} As a complementary test we examine whether the
introduction of the geographical factor improves the fit already
provided by the historical index. 

\qA{Findings}

\qun{Geographical index/separatist disturbances}

Part of the findings are summarized graphically in Fig.5. 

%
  \begin{figure}[tb]
    \centerline{\psfig{width=15cm,figure=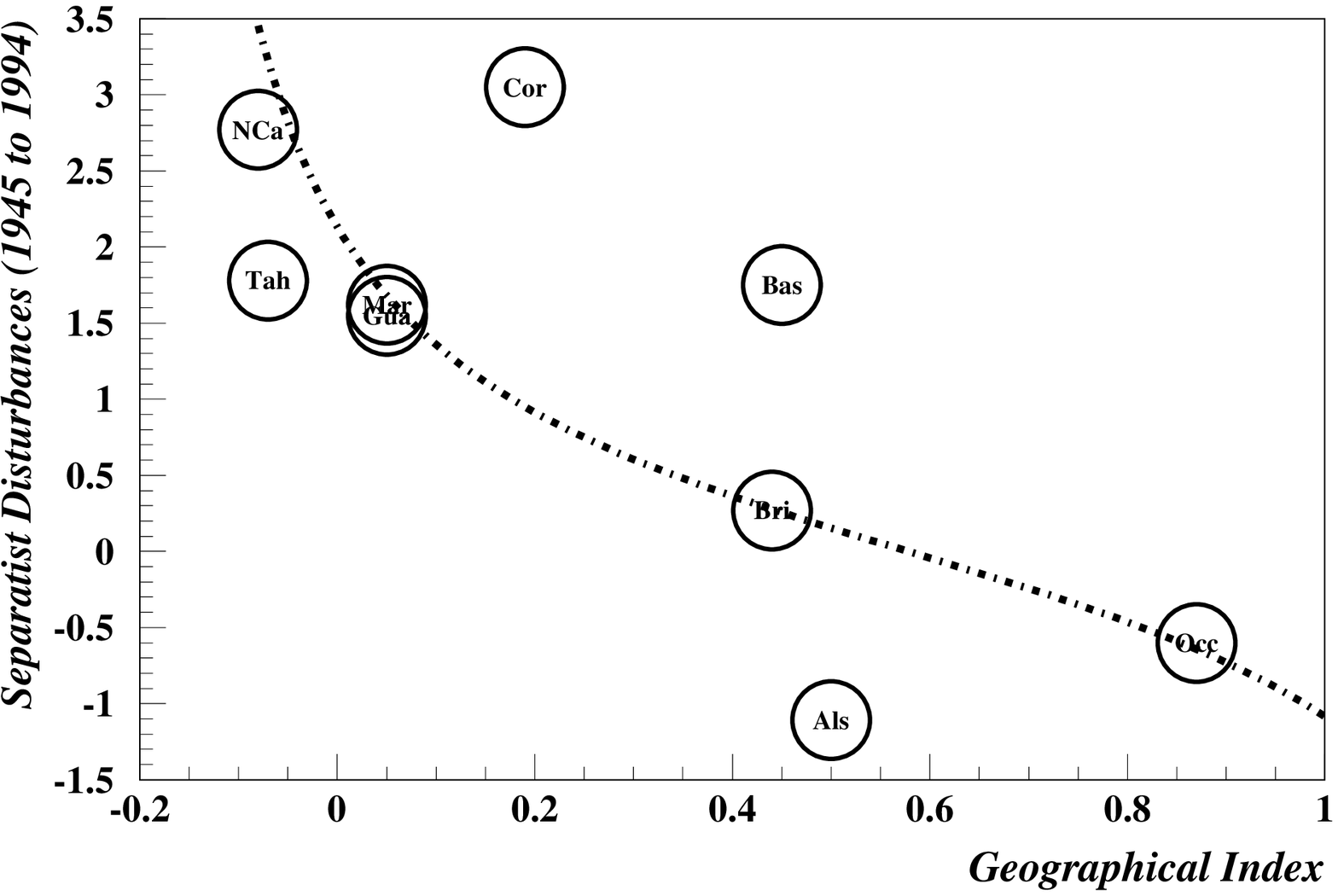}}
\qleg{Fig. 5\qhv Relationship between the geographical index and 
the level of
separatist disturbances for French provinces and Associated Territories.}
{The geographical index is an estimate of the frequency and intensity of the
interactions between the core of the country and its peripheral regions.
The curve is a least-square fit; the correlation is equal to -0.79.
The symbols have the following significations: Als: Alsace; Bas: Basque
Country; Bri: Brittany; Cor: Corsica;
Gua: Guadeloupe; Mar: Martinique; NCa: New 
Caledonia; Occ: Occitania; Tah: Tahiti. The two almost overlapping 
circles are the two close islands of Martinique and Guadeloupe.}
{}    
 \end{figure}

Fig.5 shows the relationship between the spatial index $ g $ and
the intensity of separatist disturbances as measured by the index
$ d $. The estimated equation is%
\qfoot{The fact that the estimated equation is a tangent rather
than a straight line comes from the fact that $ g $ is bounded
between $ -1 $ and $ 1 $; notice however that a linear fit is only
slightly less effective: $ r= -0.71 $}%
: 
$$ d = \tanh (-1.89g +1.1) \qquad r=-0.77 \quad (-0.22,-0.95) $$

where $ r $ denotes the coefficient of correlation; the figures in
parenthesis give the confidence interval to probability 0.95.

\qA{Influence of the historical factor}

In this section, we will see that the intensity of separatist struggles
in the second half of the twentieth century is essentially determined by their
intensity in the preceding century, 
that is, from 1845 to 1945. In other
words, separatist struggles are characterized by a strong historical
continuity. On the level of principles, there is no particular reason to
limit the historical retrospective to one century, but the difficulty of
finding sufficiently precise sources for the earlier periods imposes this
limitation. Even the period 1845--1945 remains full of uncertainties
for a number of non-European nations.
\qL
In order to estimate the intensity of separatist struggles, we will use the
number of equivalent deaths defined in chapter 6. As we will use
that index in two different time periods we need two different notations:
the number of equivalent deaths in the period 1845--1945 will be called the
{\it historical index} and denoted
by $ h $ in order to distinguish it from the same index
in the period 1945--1995 for which we keep the notation $ d $. 
\qL
For this study, we will distinguish European and non-European nations, for
at least two reasons. First of all, in the non-European countries, there
was often a conquest phase in the nineteenth century, frequently accompanied by
open warfare between the central power and the autochtonous nations. These
wars involved armed conflict on a grand scale. This characteristic is
relatively rare in Europe in the same period. In addition, in the majority
of the countries where there were such wars of conquest, the numbers of
the 
autochtonous population is rarely known with any accuracy. The dearth of
statistical sources leads us to use an index in which the number of deaths
during separatist disturbances is used straight out and is not related to
the overall population. In order to distinguish these two types of
historical indexes, we will call them $ h $
when the deaths are related to the
population and $ h' $ when they are not. For European nations, 
we then use $ h $,
while for the non-European nations we will use $ h' $.

\qun{European nations}

Figure 8.10a shows how postwar separatist troubles are determined by the
level of separatist disturbances during the preceding century. The
coefficient of correlation between the two variables is equal to 0.56 (with
a confidence interval at 95 percent given by -0.11, 0.89).
%
  \begin{figure}[p]
    \centerline{\psfig{width=14cm,figure=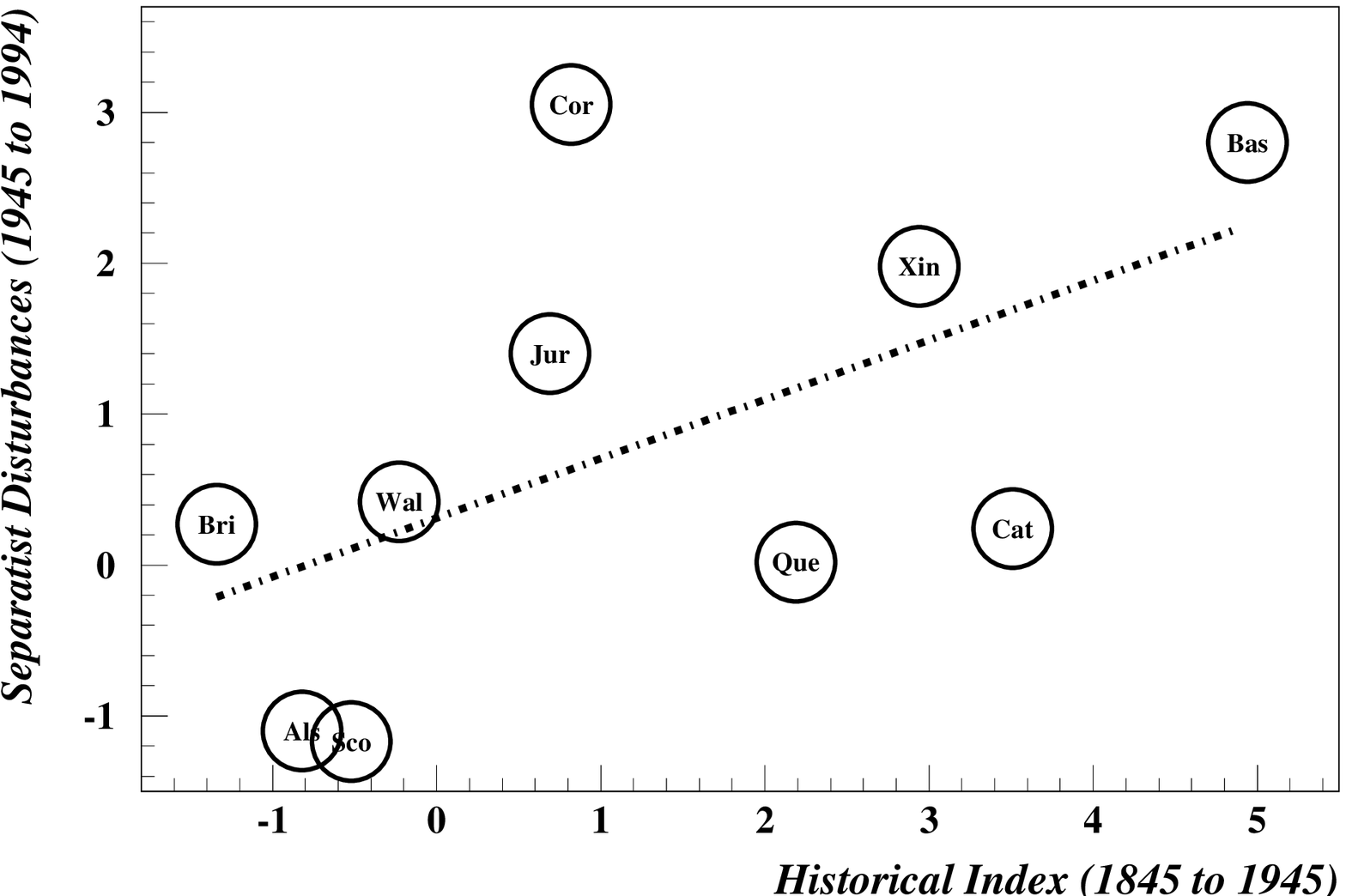}}
\qleg{Fig. 6a\qhv 
Relationship between the historical index and the level of
separatist disturbances for European nations in the period 1945--1994.} 
{The
historical index $ h $ refers to the period 1845--1945 and is given by: 
$ h=\log (\hbox{number of deaths}/$population of the minority).
The coefficient of correlation is equal to 0.56.
The labels have the following significations:
Als: Alsace; Bas: Basque Country; Bri: Brittany; Cat: Catalonia;
Cor: Corsica; Jur: Bernese Jura; Que: Quebec; Sco: Scotland;
Wal: Wales; Xin: Xinjian.}
{}
\vskip 4mm
    \centerline{\psfig{width=14cm,figure=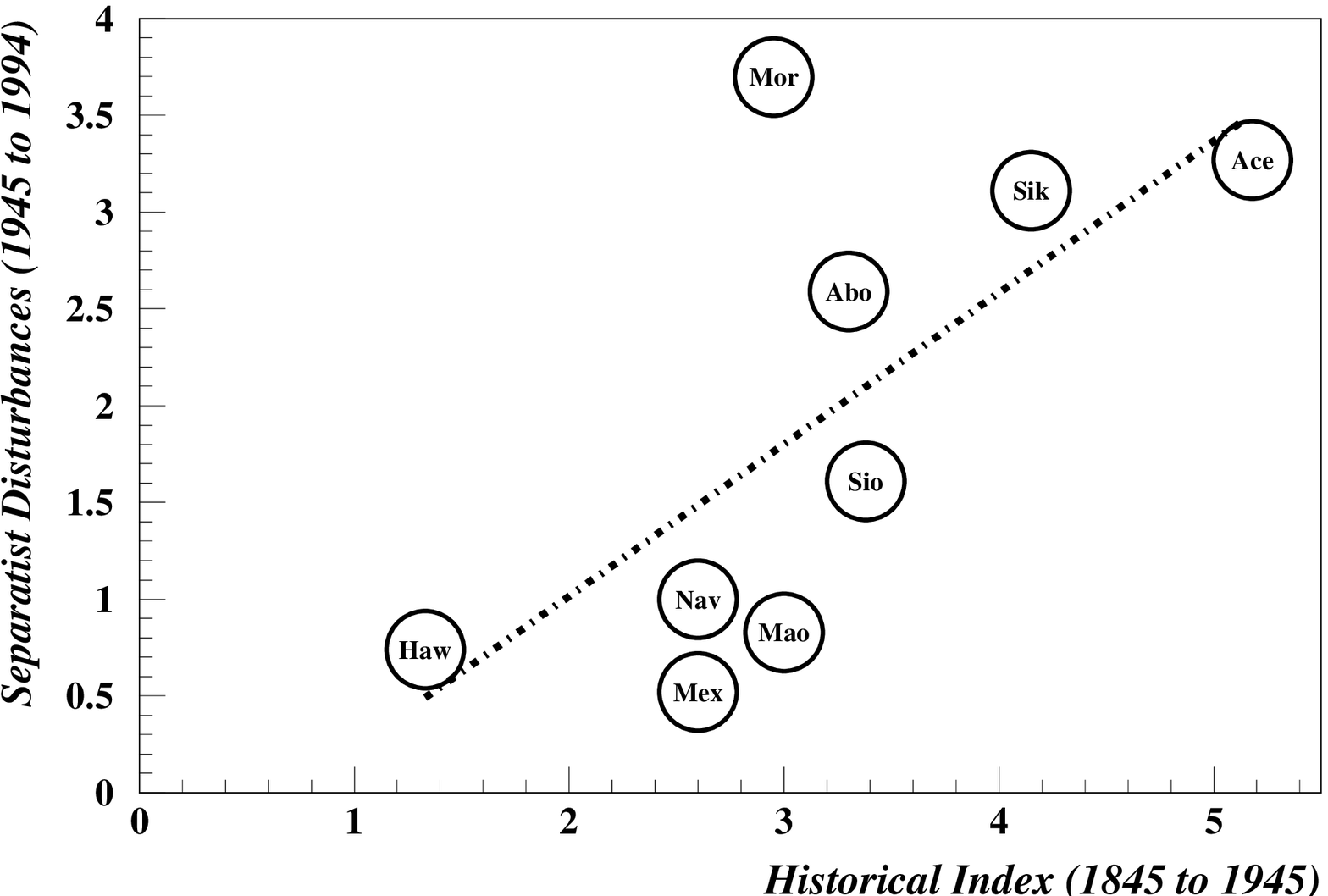}}
\qleg{Figure 6b\qhv 
Relationship between the Historical Index and the Level of
Separatist Disturbances for Non-European Nations in the Period 1945--1994.}
{The historical index $ h' $ refers to the period 1845--1945 and is given by: 
$ h'=\log (\hbox{number of deaths}) $. The
coefficient of correlation is equal to 0.68.
The labels have the following signification: Abo: Australian
Aborigenes; Ace: Aceh; Haw: Hawaii; Mao: Maoris; Mex: Mexican Americans
in New Mexico and Arizona; Mor: Moros in Mindanao and Jolo
(Philippines); Nav: Navajos; Sik: Siks in Punjab (India);
Sio: Sioux.}
{}
 \end{figure}
The regression equation reads:
$$ d=ah +b \quad a=0.39,\ b=0.32 $$

\noindent where:

\noindent $ d $: level of separatist disturbances in the period 1945--1994;

\noindent $ h $: level of separatist disturbances in the period 1845--1945;

\noindent $ h=\log (\hbox{number of deaths/population}) $

We note that since $ a $ is smaller than one the intensity of the
disturbances has on average decreased. More specifically 
the intensity of disturbances has obviously
gone up in  Corsica and the Jura, while it has gone down in
Catalonia and Quebec.
\qL
We can see in Figure 8.10a that two minorities stand out with respect to
the regression line: Catalonia, where past disturbances were too
important and Corsica, there they were too limited. These two cases
illustrate the difficulty of tracing a clear line between separatist
disturbances and other kinds of problems. In the statistics on Catalonia,
we included the Carlist wars as well as the 1936--1939 Civil War; in these
conflicts, there was indeed a defense of regional
autonomy, but this was not the only aspect. The conflicts between
monarchists and republicans or between socialists and conservatives
were intertwined with separatist struggles to the point 
where a clear separation
between them becomes impossible. In the case of Corsica, the period
1845--1945 was marked by very few openly separatist acts; nevertheless,
Corsica underwent chronic banditry during this
period (on this topic, see the book by X. Versini); insofar as this
banditry was in opposition to 
the established order represented by customs agents and other
officers, it can be considered as an form of separatist struggle. However,
it was not included in the statistics that we used.

\qun{Non-European nations}
Figure 8.10b shows the link between postwar separatist disturbances and
those that occurred during the preceding century. The coefficient of
correlation between the two variables is 0.68 
(a 95 percent confidence interval is 0.02, 0.92).
The regression equation reads:
$$ d=ah' + b \quad a=0.78,\ b=-0.55 $$

\noindent where:
\qL
\noindent $ d $: level of separatist disturbances in the period 1945--1994;
\qL
\noindent $ h' $: level of separatist disturbances in the period 1845--1945. 
\qL
In order to determine if there has been an increase or decrease in the rate
of disturbances, we can not directly compare $ d $ and $ h' $
because these two
variables are no longer defined in the same way. Qualitatively, it is
nonetheless clear that the
intensity of disturbances has greatly lessened for Aborigines, Maoris,
Mexicans, Navajos and Sioux.

\qA{Basic tendencies in the manifestations of separatist
disturbances}

In the introductory chapter of ``European Revolutions'' Charles
Tilly develops a suggestive parallel between revolutions and
traffic jams; he notes that like the latter, revolutions 
may be difficult to predict but that ``once begun, they display
recurrent patterns%
\qfoot{In the
case of traffic jams such patterns are particularly obvious:
efforts of those on the periphery to exit from the scene and
fierce competition for only small advantages for those in the
middle.}%
'';  
In this paragraph we undertake the same kind of analysis: taking
the occurrence of separatist struggles for granted we examine
if there are some permanent patterns in the variety of their
forms. The reader should be cautioned that such a study in
microsociology requires very detailed historical information which
may not necessarily be available in every case; consequently, our
results may appear somewhat partial and frustrating. In particular
the classes of separatist disturbances that we introduced in Table
B1 (Appendix B) are still somewhat too broad. Even so, Table B2
reveals two notable features. 
\qee{1} A progressive decrease in the
violence-level from the 19th to the 20th century, especially in
industrialized countries.  
\qee{2} A high level of permanence in
the forms assumed by separatist disturbances in a given country
in the course of time.  \qL
Let us now discuss these two features in some detail. The first
statement particularly applies  to the following cases: Basque
Provinces (Spain), Brittany, Corsica, Indians in the United
States, Maori (New Zealand), Quebec, Scotland, Wales. Here again,
however, a distinction should be made between European and
non-European countries. In the former (with the exception of
Spain) the decrease in the level of violence already occurred in
the period 1850-1945, whereas in the latter it took place only in
the period 1945-1994. This observation comes
into proper light if one realizes that for the period 1945-1994
there was an obvious impossibility for minority groups in
industrialized countries to wage war against the state%
\qfoot{The traditional forms of resistance being barred, these
peoples will have to shift to new ones; not surprisingly, this
is a lengthly process whose completion may take well over one
century.}%
. 
Nowadays, open warfare
can only be used in tropical or mountainous regions and against
rather weak states. The military defeat of the Sikhs in Punjab
(1983-1992) in spite of their audacity and courage was just one
other illustration of that evidence.
\qpar
Regarding  the second point one should note that for early time
periods, before 1850, only {\it major} disturbances have usually
been recorded. Moreover, some of the items listed in Table B1
(for instance those regarding electoral protests) simply did not
exist. It is difficult to avoid such bias in
analyzing and documenting non violent forms of separatist protest.
Even for those events which are on record, one does usually not 
have the detailed information that would be required; a
demonstration in Wales for instance is not identical to one in
Catalonia. For all these reasons we shall have to restrict
ourselves to only a few examples. 
\qee{1} The first example concerns the celebration of the national
day (labelled 2Ar in Table B1). Almost all national groups have
chosen a specific national day, on which huge gatherings usually 
take place. In Wales for example the National Eisteddfod (which
means session) assembles thousands of Welshmen; founded in 1450, it
has become a famous national institution. Similar instances are the
Diada Nacional (11 September) in Catalonia; the Day of the Basque
Nation (12 April), the Day of the Jurassian People (8 September),
the St Jean-Baptiste (24 June) in Quebec. In the 1970s national
days were often marked by bitter street fighting between young
demonstrators and the police. More generally the purpose of the
national day is to serve as a vital lead and to manifest
the nation's vitality.
 \qee{2} To discuss another  example of  permanent
pattern we have selected a case for which
detailed information is available, namely Brittany (France)%
\qfoot{Let us recall that another
example (Bernese Jura) has already been analyzed in paper I.}%
.
More specifically, we shall document the fact that in Breton
separatist struggles bomb attacks against French historical
memorials constitute a well established tradition. During the last
sixty years there have been at least four actions of that kind. 
\qbu 7 August 1932: destruction in Rennes of the memorial
commemorating the reunion of Brittany to France in 1532 (S\'erant
1971).
\qbu 10 October 1973: destruction of the memorial commemorating
the last public speech of general de Gaulle before his resignation
in 1969. 
\qbu 26 June 1978: destruction by a fire bomb of an aisle of the
Ch\^ateau of Versailles.
\qbu November 1993: destruction of the ancient House of Parliament
in Rennes; it was set on fire by demonstrators in doubtful
circumstances. 
\qpar

Scotland too showed great interest in historical symbols, but used
a different ``repertoire''; an example was the theft of the ``Lia
Fail'', a stone used for coronation, in December 1950. 

\qI{Conclusion}

Throughout this paper we have tried
to look at diverse and seemingly unrelated episodes as
manifestations of the same recurrent process in order to bring
some kind of provisional order into a vast and chaotic field.
Of the two determinants that we examined, namely the
spatial and historical factors, the latter appears to be
predominant in the sense that the strength of nationalist
struggles is mainly set by former episodes and local traditions.
But once the impact of this factor has been discounted, 
spatial separateness turns out to play a major role in determining
the level of separatist struggles. This factor is also strongly
correlated with the degree of language maintenance.
\qpar

{\bf An agenda for future research}\quad We are convinced that the
fundamental mechanisms  responsible for separatist
tendencies have remained basically unchanged since the beginning
of the 19th century. This claim was given some plausibility by
examining a few examples for which data were available. Yet, for
lack of quantitative historical evidence a systematic
investigation was not possible. In this respect one should mention
the remarkable work performed by C. Tilly and his collaborators
(Horn and Tilly 1986, Tilly 1992) for England (1758-1834) and also
for France. Once
 a similar task has been carried out for other
countries a systematic investigation of separatist
struggles in the 18th and 19th centuries will become possible. 
Needless to say, in the case of South-East or South-West Asian
countries this requires a working collaboration with local
scholars and librarians. 
\qpar

{\bf The future}\quad Should one expect a cardinal change in
the way people of different strands are able to communicate with
each other? In the wake of the computer revolution a substantial
change may be brought about by the introduction of a device
performing automatic speech translation. Such a process
would involve the following steps:
$$
\hbox{\it \small Language}\ A \rightarrow  \hbox{\it \small Speech
recognition}  \rightarrow  \hbox{\it \small Automatic translation}
\ A/B \rightarrow 
 \hbox{\it \small Vocal synthesis of language}\ B $$

It is of course difficult to guess how long it will take for such
a device to become widely used: 30,
75 or 100 years? Yet, even 100 years is a short span of time in
comparison to the stretches of several centuries required by the
process of linguistic assimilation. Such a translation
device would give multi-linguistic states a chance to realize their
linguistic unity much more easily that could be expected otherwise.

 \vfill \eject

\appendix
\qI{Appendix A: The data base}
\qA{Selection of the minority groups}
The minority groups documented below should be seen as a sample of
a larger set of cases. In his comprehensive survey of
minorities, T. Gurr (1993) listed 227 communal groups. As we
already pointed out in paper I, Gurr used a broader minority
concept. Our definition corresponds approximately to two of Gurr's
five subclasses, namely ``ethnonationalist groups'' and
``indigenous people''; for these subclasses Gurr's survey lists
a total of 121 groups. Our own selection of about 40 cases was
based on the following criterions. 
\qbu Data availability
\qbu Within the previous constraint we tried to make the sample
as representative as possible by selecting minority groups from
the five continents. \qL
The first criterion lead us to leave aside countries from the
ex-USSR. For the same reason, minorities in Africa and Asia are
under-represented. We also left aside those separatist struggles
which, besides their domestic significance, have become major
issues in international relations. This is for instance the case
for Ulster, for former Yugoslavia, for the Kurd minorities or for
Cyprus. 
 
\qA{Equivalence factors}
In order to build an intensity-index suitable for a great variety
of separatist movements, we converted standard separatist events
into equivalent numbers of deaths. The following factors have been
used as equivalents for one death: 
\qbu Demonstration of one million people
\qbu 30 bombs or 60 arson attempts
\qbu 30 non fatal casualties
\qbu Petition of 100 000 people \qL
Except for the last factor, which has in fact been used rather
rarely, the other equivalence factors are based on fairly
realistic orders of magnitude. Once the equivalent number of
deaths $ n $ has been obtained the intensity index is defined by
the logarithm of $ n $ referred to the total minority population $
p $: 
$$ d= \lg (n/p) $$
In taking the logarithm we follow a standard practice (see for
instance Richardson) which has two distinctive advantages. (i) it
keeps the magnitude of $ d $ of the order of a few units which is
appropriate for an index (ii) it minimizes the incidence of
estimation errors.

\qA{Sources}
Our data set is mainly based on event analysis using major
newspapers. Most valuable in that perspective has been the
newspaper data base of the ``Fondation Nationale des Sciences
Politiques'' (27 rue Saint Guillaume, Paris). It includes articles
from a variety of papers (in English, French, German, Italian and
Spanish) arranged by countries and topics of interest; the base
covers the whole period from 1945 to present. The coverage is not
uniform of course, neither in time nor in space; that kind of bias
is discussed in  Olzak (1992); by and large, however, one may
subscribe to S. Olzak's conclusion that ``few alternative sources
contain so much information''.  Anyone who has attempted a
comparative study based upon published reports not originally
prepared for this purpose has experienced the frustration that
arises when the accounts become mute just at the point where some
crucial proposition it to be tested. Needless to say, we also used
monographs and journal articles whenever they permitted to fill a
major gap. 

\vfill \eject

\qI{Appendix B: Forms of separatist disturbances}
\qA{Categorizing forms of separatist disturbances}
We introduce the following classification of separatist
manifestations.


\def\qII#1#2
{\hbox{\hbox to 5mm{\hfill #1}\hbox to
3mm{}\hbox to 11.2cm{ #2  \hfill}}}


\def\qAA#1#2
{\hbox{\hbox to 10mm{}\hbox to 10mm{\hfill #1}\hbox to 3mm{}%
\hbox to 9.7cm{  #2  \hfill}}}

\def\qunu#1#2
{\hbox{\hbox to 20mm{}\hbox to 10mm{\hfill #1}\hbox to 2mm{}%
\hbox to 8.8cm{ #2 \hfill}}}

\qII{1}{Official and legal procedures}
\qAA{1A}{Petitions addressed to the government. Demand for Home
Rule}
\qAA{1B}{Expressing separatist claims in general elections or in
(official or unofficial) referendums}
\qAA{1C}{Lawsuits, to regain lost tribal lands or to get
compensation for former abuse}
\vskip 0.8mm

\qII{2}{Demonstration}
\qAA{2A}{Demonstration in town}
\qunu{2Ar}{Annual gathering (national day)}
\qAA{2B}{March into a city or from one city to another}
\qAA{2C}{Skirmishes between the demonstrators and the police (or
counter-demonstrators)}
\qAA{2D}{Roadblocking - General strike}
\qAA{2E}{Taking control of estates, villages, public buildings,
etc}
\qAA{2F}{Military occupation - State of emergency - State of
siege}
\vskip 0.8mm

\qII{3}{Destruction of public or private property}
\qAA{3A}{Destruction of public property}
\qunu{3Aa}{by arson}
\qunu{3Ab}{by bombs}
\qAA{3B}{Destruction of private property}
\qunu{3Ba}{by arson}
\qunu{3Bb}{by bombs}
\vskip 0.8mm

\qII{4}{Murder}
\qAA{4A}{Murder of officers}
\qunu{4Ab}{bombs}
\qunu{4Ash}{shooting}
\qunu{4Ast}{stabbing}
\qAA{4B}{Murder of private citizens}
\qunu{4Bb}{bombs}
\qunu{4Bsh}{shooting}
\qunu{4Bst}{stabbing}
\vskip 0.8mm

\qII{5}{Murder of several people}
\qAA{5A}{Attack against police/army posts or patrols}
\qunu{5Ab}{bombs}
\qunu{5Ash}{shooting}
\qAA{5B}{Murder of several private citizens}
\qunu{5Bb}{car bomb, bomb in public area}
\qunu{5Bsh}{shooting}
\vskip 0.8mm

\qII{6}{Warfare}

\qpar

{Table B1\quad Classification of separatist disturbances.}
The entries are classified by order of increasing bitterness and
magnitude.

\vfill \eject

\qI{Appendix C: The definition of the spatial index for an island}
As a typical example let us examine the case of Corsica (France).
It is connected to the continent through a number of shipping
lines linking the Corsican ports of Ajaccio and Bastia to French
and Italian ports. First of all we consider  relative distances.
\qpar

\qee{1} {\bf France-Corsica}\quad
There are three main ports on the French coast: Marseilles, Nice
and Toulon. Distances to Corsican ports are as follows:
$$ \matrix{
\hbox{Bastia-Nice} \hfill & \hfill : 220 \hbox{km} & 
\hbox{Bastia-Toulon} \hfill & \hfill : 320 \hbox{km} & 
\hbox{Bastia-Marseilles} \hfill & \hfill : 340 \hbox{km} \cr 
\hbox{Ajaccio-Nice}  \hfill & \hfill : 220 \hbox{km} & 
\hbox{Ajaccio-Toulon} \hfill & \hfill : 260 \hbox{km} & 
\hbox{Ajaccio-Marseilles} \hfill & \hfill : 300 \hbox{km} \cr } $$
$$ \hbox{\bf Average: 276 km} $$
\qee{2} {\bf Italy-Corsica}\quad
There are three main ports an the Italian coast: Genoa, Livorno
and Piombino (facing the Elba island). Distances to
Corsican ports are as follows: 
$$ \matrix{
\hbox{Bastia-Genoa} \hfill & \hfill : 200 \hbox{km} &   
\hbox{Bastia-Livorno} \hfill & \hfill : 130 \hbox{km} &   
\hbox{Bastia-Piombino} \hfill & \hfill : 70 \hbox{km} \cr 
\hbox{Ajaccio-Genoa} \hfill & \hfill : 300 \hbox{km} &   
\hbox{Ajaccio-Livorno} \hfill & \hfill : 300 \hbox{km} &  
\hbox{Ajaccio-Piombino} \hfill & \hfill : 280 \hbox{km} \cr } $$
$$ \hbox{\bf Average: 216 km} $$
\qpar

Although the average distance to Italian ports is shorter than to
French ports there are approximatively three times more trips to
France than to Italy: about 700 trips/year against 200
trips/year. Therefore it is not unreasonable to use the following
weighting factors:
$$ \matrix{
\hbox{Contacts with France} \hfill &   w_+/d_+  & d_+ =276 & w_+=3
\cr 
\hbox{Contacts with Italy} \hfill &   w_-/d_-  & d_- =216 & w_+=1
\cr } $$

Next, let us consider the multiplying factor which has been
introduced in definition 2, namely: 
$$ 1/D,\quad D= \hbox{Max}[\hbox{Inf}(d_+,d_-),3] $$

where the ``effective distance'' $ D $ is expressed in hundreds of
kilometers. The role of this factor is to ensure a reasonable
degree of continuity between the case of an island and that of a
peninsula. In particular the factor of separateness $ g $ should
as a rule be smaller for an island than for a peninsula. This is
achieved through the term 3 which means that even if $ d_+ $ and
$ d_- $ are very small (which corresponds to an island separated
from the mainland by a narrow passage of water, as in the case of
the Island of Wight for instance) the effective distance 
$ D $ will be equal to 3 which accounts for the  cost
of loading and unloading operations.

\vfill \eject

{\bf References}
\vskip 1cm

\def\qparr{ \vskip 1.0mm plus 0.2mm minus 0.2mm \hangindent=10mm
\hangafter=1}

\qparr
ANTONETTI (P.) 1973: {Histoire de la Corse.} Robert Laffont.
Paris.

\qparr
AQUARONE (M.-C.) 1987: {Les fronti\`eres du refus. Six
s\'eparatismes africains.} Editions du Centre National de la
Recherche Scientifique. 

\qparr
BANGOU (H.) 1989: {La R\'evolution et l'esclavage \`a la
Guadeloupe.} Messidor-Editions Sociales. Paris. 

\qparr
BARR\`ES DU MOLARD (A.) 1842: {M\'emoires sur la guerre de la
Navarre et des Provinces basques depuis son origine en 1833
jusqu'au trait\'e de Bergara en 1839.} Dentu. Paris. 

\qparr
BARROS (J.) 1968: {The Aland Islands question: its settlement
by the League of Nations.} Yale University Press. New Haven.

\qparr
BASSAND (M.) et al 1985: {Les Suisses entre la mobilit\'e et
la s\'edentarit\'e.} Presses Polytechniques Romandes. 

\qparr
BENJAMIN (T.) 1989: {A rich land, a poor people. Politics and
society in modern Chiapas.} University of New Mexico Press.
Albuquerque. 

\qparr
BIRCH (A.H.) 1977: {Political integration and disintegration
in the British Isles.} George Allen and Unwin. London.

\qparr
BRUN (A.) 1946: {Parlers r\'egionaux. France dialectale et
unit\'e fran\c caise.} Didier. Paris.

\qparr
BURTON (R.D.) 1994: {La famille coloniale. La Martinique et
la m\`ere patrie 1789-1992.} L'Harmattan. Paris. 

\qparr
CLOUGH (S.B.) 1930: {A history of the Flemish movement in
Belgium. A study in nationalism.} Richard R. Smith. New York.

\qparr
COOK (S.F.) 1976: {The population of the California Indians
1769-1970.} University of California Press.

\qparr
COUPLAND (R.) 1954: {Welsh and Scottish nationalism.} Collins.
London. 

\qparr
CRIBB (R.) 1992: {Historical dictionary of Indonesia.} The
Scarecrow Press. Metuchen (N.J.).

\qparr
DAVANT (J.L.) 1975: {Histoire du Pays Basque.} Editions
Elkar. Bayonne. 

\qparr
DAVIES (D.G.) 1970: The concentration process and the growing
importance of non central governments in federal states. Public
Policy 18,5,649-657.

\qparr
DENIER (J.) 1919: {L'attribution des Iles d'Aland.} Paris.

\qparr
DENNIS (H.C.) ed 1977: {The American Indians 1492-1976. A
chronology and fact book.} Oceana Publications. 

\qparr
DOUSSET-LEENHARDT (R.) 1976: {Terre natale, terre d'exil.}
Maisonneuve et Larose. Paris.

\qparr
DUPUY (R.) 1995: La participation anglaise au d\'ebarquement de
Carnac (26 juin 1795). Histoire,June,66-71.

\qparr
DWELSHAUVERS (G.) 1926: {La Catalogne et le probl\`eme
catalan.} F\'elix Alcan. Paris.

\qparr
FISTI\'E (P.) 1985: {La Birmanie ou la qu\^ete de l'unit\'e.
Le probl\`eme de la coh\'esion nationale dans la Birmanie
contemporaine et sa perspective historique.} Publications de
l'Ecole Fran\c caise d'Extr\^eme Orient. Vol.79.

\qparr
FLORA (P.) 1983: {State, economy and society in Western
Europe 1815-1975. A data handbook.} Vol.I. Campus Verlag,
Macmillan Press, St James Press, Frankfurt.

\qparr
GILBERT (M.) 1968: {British history atlas.} Weidenfeld and
Nicholson. London.

\qparr
GURR (T.R.) 1993: Why minorities rebel: a global analysis of
communal mobilization and conflict since 1945. International
Political Science Review 14,2,161-201.

\qparr
HECHTER (M.) 1975: {Internal colonialism. The Celtic fringe
in the British national development 1536-1966.} Routledge and
Kegan Paul. London.

\qparr
H\'ERAND (F.) 1993:  L'unification linguistique de la France.
Population et Soci\'et\'es 285,Dec,1-4.

\qparr
HISTOIRE des provinces de France 1984: Fernand Nathan. Paris.

\qparr
HORN (N.), TILLY (C.) 1986: Catalogs of contention in Britain,
1758-1834. New School for Social Research. Working Paper No32.

\qparr
HUGHES (R.) 1987: {The fatal shore.} Collins Harvill. London.

\qparr
INDEX DU TEMPS 1966: {Tables du journal ``Le Temps'':
1861-1900.} Institut Fran\c cais de Presse. Paris.

\qparr
JANSSON (J.-M.) 1961: Bi-lingualism in Finland. Fifth World
Congress of the International Political Science Association
(September 26-30).

\qparr
JENKINS (J.R.G.) 1986: {Jura separatism in Switzerland.}
Clarendon Press. Oxford.

\qparr
KIRK (D.) 1946: {Europe's population in the interwar period.}
Gorgon and Breach. New York.

\qparr
KROEBER (T.) 1968: {Ishi, testament du dernier indien sauvage
de l'Am\'erique du Nord. Plon. Paris.}

\qparr
LA COLLABORATION des autonomistes 1946: Secr\'etariat d'Etat \`a
la Pr\'esidence du Conseil (26 mai).

LA M\'ELAN\'ESIE 1946: Documentation fran\c caise. 

\qparr
LERUEZ (J.) 1983: {L'Ecosse, une nation sans Etat.} Presses
Universitares de Lille. Lille.

\qparr
LETAMENDIA (P.) 1987: {Nationalismes au Pays basque.} Presses
Universitaires de Bordeaux.

\qparr
LIEBERSON (S.), HANSEN (L.K.) 1974: National development, mother
tongue diversity and the comparative study of nations. American
Sociological Review 39,523-541.

\qparr
LIEBERSON (S.), DALTO (G.), JOHNSTON (M.E.) 1975: The course of
mother-tongue diversity in nations. American Journal of Sociology
81,1,34-61

\qparr
LIEBERSON (S.) 1985: {Making it count. The improvement of
social research and theory.} University of California Press.

\qparr
MEIER (M.S.), RIVERA (F.) 1972: {The Chicanos, a history of
Mexican Americans.} Hill and Wang. New York.

\qparr
M\'EMOIRE 1920: La question des Iles d'Aland et le droit de la
Finlande. M\'emoire pr\'esent\'e par un groupe de juristes et
d'historiens finlandais. Helsinki.

\qparr
OLIVER (M.) 1961: Political problems of poly-ethnic countries:
Canada. 5th Word Congress of the International Political Science
Association. 

\qparr
OLZAK (S.) 1992: {The dynamics of ethnic competition and
conflict.} Stanford University Press. 

\qparr
POMPONI (F.) 1979: {Histoire de la Corse.} Hachette. Paris.

\qparr
PRYOR (F.L.) 1968: {Public expenditures in communist and
capitalist nations.} George Allen and Unwin. London.

\qparr
REBELLION in the Canadas (1837-1838) 1987: Ottawa.

\qparr
RENNWALD (J.C.) 1984: {La question jurassienne.} Editions
Entente. Paris.

\qparr
REUSS (R.) 1912,1916: {Histoire d'Alsace.} Boivin et Cie.
Paris.

\qparr 
ROEHNER (B.M.) 1997: Spatial and historical determinants
of separatism and integration.
Swiss Journal of Sociology 23,1,25-59.\qL
[This paper is available online in two parts at the following
addresses:\qL
http://www.lpthe.jussieu.fr/~roehner/$ \sim $sep1.pdf (part 1)\qL
http://www.lpthe.jussieu.fr/~roehner/$ \sim $sep2.pdf (part 2)]

\qparr
ROEHNER (B.M.) 2002: Separatism and integration.
A study in analytical history. Rowman and Littlefield, Laham
(Maryland).

\qparr
ROEHNER (B.M.) 2017: Separatism and disintegration.\qL
[This revised and updated version of the previous publication
is available online at the following address:\qL
http://www.lpthe.jussieu.fr/$ \sim $roehner/separatism.pdf]

\qparr
ROSSINYOL (J.) 1974: {Le probl\`eme national catalan.}
Mouton. Paris. 

\qparr
RICHARDSON (L.F.)  1960: {Statistics of deadly quarrels.}
Boxwood Press. Pittsburgh.

\qparr
SCHINDLER (D.) 1959: Entwicklungstendenzen des Schweizerischen
F\"oderalismus. Schweizer Monatshefte 8,697-709.

\qparr
S\'ERANT (P.) 1971: {La Bretagne et la France.} Fayard. Paris.

\qparr
SINGLETON (F.) 1989: {A short history of Finland.} Cambridge
University Press. Cambridge.

\qparr
SOROKIN (P.A.) 1937: {Social and cultural dynamics. Vol.III:
Fluctuations and social relationship. War and revolution.}
American Book Company.

\qparr
TERRA (R.) 1993: Le romanche, quatri\`eme langue des Suisses.
Annales de G\'eographie 574,596-610.

\qparr
TILLY (C.) 1992: How to detect, describe and explain repertories
of contention. New School for Social Research. Working Paper No150.

\qparr
TILLY (C.) 1993: {European revolutions 1492-1992.} Oxford
University Press. Oxford.

\qparr
TIMM (H.) 1961: Das Gesetz der wachsenden Staatsausgaben.
Finanzarchiv 21,2,201-247.

\qparr
WAGNER (A.) 1893: {Grundlegung der allgemeinen oder
theoretischen Volkswirtschaftslehre.} Leipzig.

\qparr
WOLFF (L.) 1991: {Little Brown Brother. How the United States
purchased and pacified the Philippines.} Oxford University Press.
Singapore.

\qparr
ZAINU'DDIN (A.G.T.) 1968,1980: {A short history of Indonesia.}
The Scarecrow Press. Metuchen (N.J.)

\qparr
ZELLER (G.) 1945: {L'Alsace fran\c caise de Louis XIV \`a nos
jours.} Armand Colin. Paris.

\end{document}